\documentclass[enlgish, twocolumn]{article}

\usepackage{graphicx}
\usepackage{hyperref}
\usepackage{authblk} 
\usepackage{float}
\usepackage{bbold}
\usepackage{geometry}
\usepackage{multicol}
\usepackage[square,numbers,sort&compress]{natbib}

\usepackage{amsmath}
\usepackage{bm}
\usepackage{braket}
\usepackage[utf8]{inputenc}
\usepackage{enumerate}
\usepackage{physics}
\usepackage{color,soul}
\usepackage{mleftright}
\newcommand{\subsubsubsection}[1]{\paragraph{#1}\mbox{}\\}
\usepackage{footmisc}
\usepackage{bigfoot}
\begin{document}

\title{Quantum-secure multiparty deep learning}

\author[1]{Kfir Sulimany\thanks{Corresponding author: \texttt{Kfir@mit.edu}}}
\author[1]{Sri Krishna Vadlamani}
\author[1,2]{Ryan Hamerly}
\author[1]{Prahlad Iyengar}
\author[1]{Dirk Englund}
\affil[1]{Research Laboratory of Electronics, Massachusetts Institute of Technology, Cambridge, MA, USA}
\affil[2]{Physics \& Informatics Laboratories, NTT Research, Inc., Sunnyvale, CA, USA}
\date{}
\vspace{-3em} 

\twocolumn[\begin{@twocolumnfalse}
\maketitle
\vspace{-2.5em} 
\begin{abstract}
Secure multiparty computation enables the joint evaluation of multivariate functions across distributed users while ensuring the privacy of their local inputs. This field has become increasingly urgent due to the exploding demand for computationally intensive deep learning inference. These computations are typically offloaded to cloud computing servers, leading to vulnerabilities that can compromise the security of the clients' data. To solve this problem, we introduce a linear algebra engine that leverages the quantum nature of light for information-theoretically secure multiparty computation using only conventional telecommunication components. We apply this linear algebra engine to deep learning and derive rigorous upper bounds on the information leakage of both the deep neural network weights and the client's data via the Holevo and the Cramér-Rao bounds, respectively. Applied to the MNIST classification task, we obtain test accuracies exceeding $96\%$ while leaking less than $0.1$ bits per weight symbol and $0.01$ bits per data symbol. This weight leakage is an order of magnitude below the minimum bit precision required for accurate deep learning using state-of-the-art quantization techniques. Our work lays the foundation for practical quantum-secure computation and unlocks secure cloud deep learning as a field.
\vspace{2mm} 
\end{abstract}
\end{@twocolumnfalse}]
\renewcommand{\thefootnote}{\fnsymbol{footnote}}
\footnotetext[1]{Corresponding author: \texttt{Kfir@mit.edu}}

\section{Introduction}
Although deep learning has revolutionized multiple domains \cite{lecun2015deep,vinyals2019grandmaster,brown2020language}, its applications are constrained by the increasing computational demands on the hardware \cite{li2017multi,satyanarayanan2017emergence,butt2020review}. Given the high energy consumption required for state-of-the-art deep neural network (DNN) inference \cite{sze2017efficient, yang2017method, patterson2021carbon}, it is common to delegate inference workloads from the edge to centralized server clusters. Unfortunately, this paradigm introduces vulnerabilities that compromise data security, which is crucial in applications such as business, finance, and healthcare \cite{li2017multi}. This situation underscores the central challenge in secure computation,  where multiple parties perform a joint evaluation of multivariate functions across distributed resources while preserving the privacy of their inputs \cite{yao1982protocols} (Fig.~\ref{Fig_Ilus}(a)).

Modern secure computation schemes are built on homomorphic encryption, which allows universal computation on encrypted data and preserves the security of the input and output of the computation \cite{gentry2009fully}. Recent developments have adapted state-of-the-art homomorphic encryption schemes for secure machine learning \cite{cheon2017homomorphic}, but real-world applications are limited due to their massive computational overhead \cite{butt2020review} and recently discovered security vulnerabilities \cite{li2021security}. Moreover, these encryption schemes typically depend on computational complexity and are not information-theoretically secure.

In this work, we introduce a linear algebra engine for information-theoretically secure computation. We apply our engine to multiparty deep learning and derive a rigorous security analysis for all parties. Applied to the MNIST classification task, we obtain test accuracies of more than $96\%$  while leaking less than $0.1$ bits per weight symbol and $0.01$ bits per data symbol. This weight leakage is an order of magnitude below the state-of-the-art minimum bit precision required for deep learning \cite{gupta2015deep}.

Our protocol relies on a photonic computation architecture to realize efficient optical matrix-vector multiplication \cite{wetzstein2020inference, shastri2021photonics, mcmahon2023physics,choi2024photonic}. State-of-the-art optical DNNs have been demonstrated using integrated photonics \cite{shen2017deep,tait2017neuromorphic,ashtiani2022chip,feldmann2019all,feldmann2021parallel,chen2023deep,bandyopadhyay2022single, wu2023lithography, pai2023experimentally,basani2024all}, free-space optics \cite{lin2018all,zhou2021large,wright2022deep,wang2022optical,bernstein2023single, wang2023image, ma2023quantum} and fiber optics \cite{xu202111, davis2022frequency}. Recently, a delocalized optical DNN using low-power devices was introduced and demonstrated \cite{hamerly2019large, sludds2022delocalized}. This protocol, which operates in an edge computing setting \cite{satyanarayanan2017emergence} where the DNN weights are streamed from a central server to clients that perform inference, shields client data from exposure but reveals the weights to the client. 

To address this asymmetry, the client in our protocol performs the inference and then returns the residual light to the server as a verification state for security. Leveraging the quantum nature of light, we show that the server can compute an upper-bound on their weight leakage using the Holevo theorem \cite{holevo1973bounds} and that the client can upper-bound their data leakage using the Cramér–Rao bound \cite{rao1992information}.

\section{Coherent linear algebra engine}\label{Sec_engine}
\begin{figure}[!ht] 
\begin{centering}
\includegraphics[width=\columnwidth]{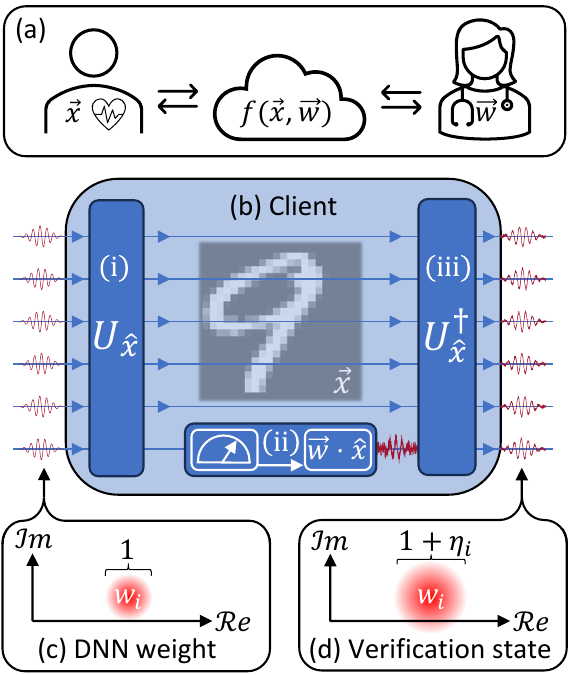}
\par\end{centering}
\caption{ \textbf{Coherent Linear Algebra Engine.} (a) Secure multiparty computation enables joint evaluation of a function $f(\vec{x},\vec{w})$ across distributed users while ensuring the privacy of the local inputs $\vec{x},\vec{w}$. (c) Deep neural network (DNN) weights are encoded into the complex amplitudes of coherent states $w_i$ (of variance $1 \text{ SNU}$ (Shot-Noise-Units)). (b) The client uses these weights for inference with their local data $\vec{x}$ and transmits the residual light, called the verification state, back to the server. (d) The verification state has the same mean as the original DNN weight $w_i$ but a higher variance of $(1+\eta_i) \text{ SNU}$, where $\eta_i$ is the excess noise.
(b) The client receives the coherent states that encode the weights (left) and calculates the inner product $\vec{w}\cdot\vec{x}$ using: (i) unitary transformation $U_{\hat{x}}$, (ii) measurement and feedforward of the complex amplitude $\vec{w}\cdot\hat{x}$, which adds noise (marked in red) to the last mode, 
 (iii) and the unitary $U_{\hat{x}}^{\dagger}$. After applying $U_{\hat{x}}^{\dagger}$, the excess noise from the measure and feedforward step is spread over the output modes (right). }\label{Fig_Ilus}
 \end{figure}

Our protocol is based on a coherent optical linear algebra engine utilized by two parties in a network: the ``server", which possesses a matrix $\mathbf{W}$ of dimensions $M \times N$ representing a DNN linear layer, and the ``client", which holds the data vector $\vec{x}$ of length $N$. The matrix-vector product $\mathbf{W} \vec{x}$ is computed through $M$ inner products $\mathbf{W}_i \vec{x}$ for $i \in \{1, \dots, M\}$, each of which involves three steps:

\begin{enumerate}
  \item The server transmits a logical weight vector \(\mathbf{W}_i\) encoded into the complex amplitudes of coherent states \(\vec{w}\). The amplitudes of these physical states, \(w_j\), encode the logical weights as \(w_j = \sqrt{\mu} \mathbf{W}_{ij} / \|\mathbf{W}\|_{\text{RMS}}\), where \(\|\mathbf{W}\|_{\text{RMS}}\) denotes the root mean square of the elements in \(\mathbf{W}\). This encoding ensures that the average photon number is \(\mu\), a value chosen by the server. These coherent states have a variance of 1 shot noise unit (SNU) in each quadrature, as illustrated in Fig.~\ref{Fig_Ilus}(c).
  \item The client operates on these amplitudes as follows, Fig.~\ref{Fig_Ilus} (b):
         \begin{enumerate}[(i)]
         \item The client calculates the inner product of the incoming weight vector $\vec{w}$ with its local input data vector $\vec{x}$ using a unitary transformation $U_{\hat{x}}$ that is designed to yield $\vec{w}\cdot\hat{x}$ into one mode of $U_{\hat{x}}\vec{w}$, where $\hat{x}$ is obtained by the $\ell^2$ normalization of $\vec{x}$. We call this mode the ``result" mode. 
         \item The client diverts the result mode, and measures both quadratures to obtain $\vec{w}\cdot\hat{x}$. Then, the client performs a feed-forward protocol by reinjecting light with the same quadratures back into the result mode. The feed-forwarded state preserves the measured state's expectation value $\vec{w}\cdot\hat{x}$ but, due to quantum noise, consists of additional Gaussian noise of variance $1$~SNU. The output of this measure and feed-forward step is labeled $\mathcal{M}(U_{\hat{x}}\vec{w})$. 
         \item The client finally applies the unitary transformation $U_{\hat{x}}^{\dagger}$ on $\mathcal{M}(U_{\hat{x}}\vec{w})$ and returns $\rho_v=U_{\hat{x}}^{\dagger}\mathcal{M}(U_{\hat{x}}\vec{w})$ as a verification state to the server for security checks. By the definition of $\mathcal{M}(U_{\hat{x}}\vec{w})$ from the previous point, the first moment of $U_{\hat{x}}^{\dagger}\mathcal{M}(U_{\hat{x}}\vec{w})$ is $\vec{w}$, while the variance of each mode $i$ has increased by $\eta_i$, see Fig.~\ref{Fig_Ilus}(d).
       \end{enumerate}
  \item Finally, the server measures the variance of each mode $i$ of the verification state in both quadratures, $(1+\eta_i)$ SNU, and calculates an upper bound for the leakage of each weight $w_i$ to the client. Note that before this security check, the verification state can be re-used by other clients to perform their own inference as discussed in Section \ref{Sec_secure_classiffication}. 
\end{enumerate}

The measure and feed-forward operation $\mathcal{M}$ outlined above can be generalized to an operation $\mathcal{G}$ that consists of a combination of phase-insensitive amplification $G$ and a beam splitter with a split ratio $1-\frac{1}{G}:\frac{1}{G}$; the former approach is obtained from the latter in the limit $G\gg1$ (see Section \ref{Methods_opt}) \cite{ralph1999all}. In this general case, the client applies a gain $G$ to the component $\vec{w}\cdot\hat{x}$, passes it through a beam splitter with the above splitting ratio, and uses the output of the $1-\frac{1}{G}$ port for its inner product computation readout while reinjecting the output of the $\frac{1}{G}$ port back into the result mode that originally carried $\vec{w}\cdot\hat{x}$. Note that in order to obtain the desired inner product, the client scales the result digitally by $\frac{\|\vec x\|}{\sqrt{G-1}}\frac{\|\mathbf{W}\|_{\text{RMS}}}{\sqrt{\mu}}$. However, for the purpose of classification tasks, the scaling step for the client is superfluous \ref{Methods_class}. 

The gain $G$ contributes amplification noise to the result mode. We calculate the variance of this mode in Section ~\ref{Methods_opt}. This variance is larger than the standard shot noise by a quantity we refer to as the ``excess noise". This excess noise is then spread via the operation of $U^\dagger_{\hat x}$ in the verification states modes, such that the excess noise in the $ith$ mode of the verification state, $\eta_i$, is weighted by $\abs{\hat x_i}^2$:

\begin{equation} \label{eq_noise}
\eta_i = \left(2-\frac{2}{G}\right) \abs{\hat x_i}^2    
\end{equation} 
where $\hat x_i$ is the $i$th element of the normalized data vector $\hat x$. 

The signal-to-noise ratio (SNR) in the client's measurement after applying the amplification and beam splitting is $\text{SNR}= \frac{G(G-1)}{2G^2-3G+2} |\vec{w}\cdot \hat{x}|^2$ (see Section~\ref{Methods_opt}). Since $\abs{\vec{w}}^2$ is proportional to the average photon occupation number $\mu$, the SNR is proportional to $\mu$.

The gain $G$, controlled by the client, and the average weight occupation $\mu$, controlled by the server, are the only hardware configuration parameters in the system. These parameters dictate the verification state excess noise, Eq.~\eqref{eq_noise}, and the SNR of the inner product measurement. In the large gain limit $G\gg1$, the excess noise, $\eta_i$, and the SNR reduce to $2x_i^2$ and $\frac{1}{2}|\vec{w} \cdot \hat{x}|^2$, respectively. This result aligns with the measure and feedforward special case. In the small gain limit $G\approx1$, the excess noise $\eta_i$ and the SNR reduce to $2(G-1)x_i^2$ and $(G-1)|\vec{w} \cdot \hat{x}|^2$, respectively. Reducing the gain disturbs the verification state less, but also reduces the SNR of the inner product obtained by the client. Therefore, the client's actions can be described by the theory of weak measurements \cite{oreshkov2005weak, lundeen2012procedure,kim2012protecting,zhang2015precision}. More details of the amplification-and-split procedure and the hardware implementation of the protocol using standard telecommunication components are presented in Section \ref{Methods_opt}.

\section{Classification accuracy}

In this section, we calculate the classification accuracy of a secure neural network that uses our coherent linear algebra engine on the standard MNIST classification task. To this end, we first trained a digital noiseless neural network on the MNIST dataset and obtained a classification accuracy of $98\%$. Then we fed the trained weights into a PyTorch model of our optical architecture to evaluate the test accuracy of our secure optical neural network, see Section \ref{Methods_class} for more details. The complete open-source package used in this work is available online \cite{QSDL}. 

We calculate the classification accuracy as a function of the average photon number occupation $\mu$ and the amplification gain $G$, see Section \ref{Methods_class}. We obtain more than $96\%$ classification accuracy with an average photon number of less than $\mu=4$ per weight and amplification gain of $G=3$. 

In contrast to previous studies in optical machine learning \cite{sludds2022delocalized, wang2022optical}, we do not measure all of the light sent to the client but only the portion corresponding to $|(\vec{w}\cdot \hat{x})|^2$ and yet obtain similar high accuracies. This fundamentally follows from the fact that the SNR in the inner product is similar whether one optically routes the inner product amplitude to one mode and measures it vs. measuring all the modes and calculating the inner product digitally. 

\section{Security analysis} \label{Sec_security}
\subsection{Weights leakage}
In this section, we use the verification state $\rho_v$ returned to the server by the client to calculate an upper bound on the amount of information about $\vec{w}$, in terms of number of bits, that could be learned by an arbitrarily dishonest client. A dishonest client tries to learn the weights using arbitrary operations other than the operations defined by our protocol. 

Here, we assume that dishonest operations are independent and identically distributed over all the incoming modes of $\vec{w}$; the client employs separable ancilla states that interact individually with each weight mode. The ancillary states are stored in a quantum memory until the end of the attack and subsequently measured independently from one another to reveal information about $\vec{w}$. Our security analysis for the weight leakage relies on results for quantum encryption \cite{scarani2009security,xu2020secure,laudenbach2018continuous,pirandola2020advances,sulimany2021high, bash2015quantum} and proceeds through the Holevo theorem \cite{holevo1973bounds}; see Section \ref{security_weights} for the complete analysis. 

We present here the final mathematical result for the weight leakage. Representing the weight occupation by $\mu$ and defining the quantities:
\begin{alignat*}{2}
    &a:= 2\mu +1 \quad &&b:= 2\mu+1+\eta_i\\
    &c:= \sqrt{4\mu^2 + 2\mu +1} \quad &&z:= \sqrt{(a + b)^2 - 4c^2},
\end{alignat*}
the weight leakage $I_{w_i}$ is bounded by the Holevo theorem:
\begin{equation} \label{eq:wleakage}
    I_{w_i}\leq g(\nu_1)+g(\nu_2)-g(\nu_3)
\end{equation} 
where
\begin{equation*}
\begin{gathered}
g(\nu) = \left( \frac{\nu + 1}{2} \right) \log_2 \left( \frac{\nu + 1}{2} \right) - \left( \frac{\nu - 1}{2} \right) \log_2 \left( \frac{\nu - 1}{2} \right) \\
\nu_{1,2} = \frac{1}{2} \left( z \pm \left[ b - a \right] \right),\ \ \nu_{3} = a - \frac{c^2}{b + 1}
\end{gathered}
\end{equation*}
This inequality is shown to be tight by analyzing the entangling cloner attack \cite{laudenbach2018continuous}. The derivation of Eq.~\eqref{eq:wleakage}, and more details, can be found in Section \ref{security_weights}.

The leakage is calculated for one query, that is, a single exposure of the client to the weights. To prevent information leakage after multiple uses of the model, the server sends the client invariants of the DNN model under affine transformations consisting of different individual weights while parameterizing the same model function and, therefore, conserving the inference output; see \ref{security_multiquery}.

\subsection{Data leakage}
To calculate the data leakage, we consider an honest client and a dishonest server. In this setting, the server sends a weight matrix $\mathbf{W}$ of shape $M\times N$ to the client and aims to evaluate the client's data vector $\vec{x}$ of length $N$ using $M$ measurements of the verification state.

We recall from Section 2 that the mode $i$ of the verification state $U_{\vec{x}}^{\dagger}\mathcal{G}(U_{\vec{x}}\vec{w})$ has the mean $w_i$ and variance $(1+\eta_i$ \text{ SNU}), Eq.~\eqref{eq_noise}. In other words, information about the client's data $\vec{x}$ is leaked to the server not through the mean of the verification state but through its variance. The client may control this leakage by reducing the gain $G$ in its amplification-and-splitting step. We also note that since the client is honest, it announces the true gain $G$ to the server. 

We phrase dishonest attempts of the server to learn $\vec{x}$ through $M$ measurements as the following concrete statistical estimation problem\cite{polino2020photonic}: Let $\rho_v$, the state sent back to the server by the client, be a Gaussian random vector with mean $\vec{w}$ and a covariance matrix whose $i$-th diagonal element is $\sigma_i^2=1+\eta_i=1+(2-\frac{2}{G})\cdot |\hat x_i|^2$. What is the amount of information, in terms of bits, that the server can obtain about $\vec{x}$ from $M$ independent measurements of $\rho_v$, given that it knows $G$? 

We answer this question by using the Cramér–Rao inequality \cite{rao1992information} to compute a lower bound on the variance $\sigma_{\hat{X}_i}$ of the server's estimator $\hat{X}_i$ of the true $\hat x_i$: $\sigma_{\hat{X}_i} \geq \frac{1}{M\cdot I(\hat x_i)}$, where $I(\hat x_i)$ is the Fisher information of the true normalized data element $\hat x_i$.

Next, we assume that the server's estimator $\hat{X}_i$ is a Gaussian random variable with mean $\hat x_i$ and variance given the above lower bound and use the formula for the communication capacity of a Gaussian channel to calculate the amount of mutual information $I_{x_i}$, in number of bits, between the true unnormalized $x_i$ and the estimator $\hat{X}_i$. By the data processing inequality, the information $I_{x_i}$ leaked about $x_i$ is upper-bounded by the information $I_{\hat x_i}$ leaked about $\hat x_i$. In the case where the server has access to quantum operations, we use the quantum Cramér–Rao inequality to obtain the leakage bound. We present the derivation for the classical and quantum cases in Section \ref{security_data} and Section~\ref{security_quantum_data}, respectively. The resulting bound for the information leakage $I_{x_i}$ is:
\begin{align} \label{eq:xleakage}
    I_{x_i} &\leq \frac{1}{2} \log_2\left(1 + k\cdot 
 \frac{8 M (G-1)^2 |\hat x_i|^4}{G^2 \sigma_i^4}\right)
\end{align}
\[
    \begin{cases}
        k = 1 \implies \text{classical Cramér-Rao bound}\\
        k = 2 \implies \text{quantum Cramér-Rao bound}
    \end{cases}
\]
where $k=1$ and $k=2$ denote the classical and quantum operations, respectively, used by the server. The data leakage is zero for a gain value of $G=1$, increases with the gain, and asymptotically obtains the leakage for direct measurement and feed-forward.

\section{Secure classification} \label{Sec_secure_classiffication}

\begin{figure}[ht!]
\begin{centering}
\includegraphics[width=\columnwidth]{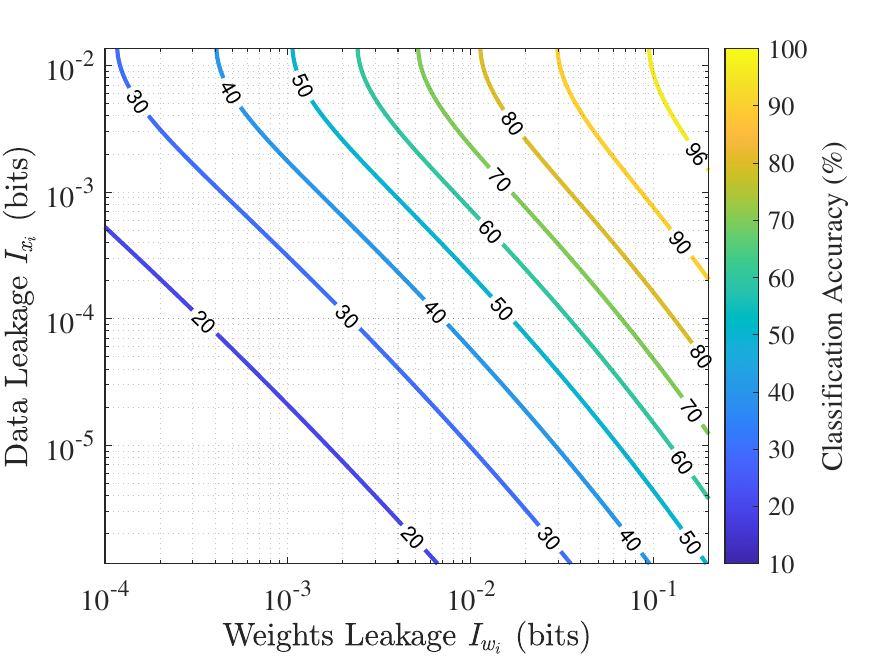}
\par\end{centering}
\caption{\textbf{Classification accuracy versus weights leakage and data leakage.} The classification accuracy of the secure optical neural network, which uses our coherent linear algebra engine (Fig.~\ref{Fig_Ilus}), is first numerically calculated as a function of the average photon occupation per weight and the amplification gain. We also use these parameters to upper bound the weight leakage $I_{w_i}$ (Eq.~\eqref{eq:wleakage}) and the data leakage $I_{x_i}$ (Eq.~\eqref{eq:xleakage}). This chain of reasoning enables the presentation of the classification accuracy as a function of the weights and data leakages. The classification accuracy increases with both leakages and achieves the digital noiseless accuracy. For small data leakages, the weight leakage is inversely proportional to the data leakage for any given fixed classification accuracy. Our protocol achieves a classification accuracy of $>96\%$ while leaking less than $I_{w_i}=0.1$ bits per weight symbol less than $I_{x_i}=0.01$ bits per data symbol. This level of bit precision in the weights is currently known to be insufficient for achieving high accuracy in state-of-the-art deep neural networks.}
    \label{Fig_main_maps}
 \end{figure}

Now that we have a mathematical handle on both the weight (Eq.~\ref{eq:wleakage}) and the data leakage (Eq.~\ref{eq:xleakage}), we can ask how they trade off against each other if one requires a constant test accuracy. For this purpose, we revisit the classification accuracy that we computed numerically as a function of the two hardware configuration parameters of the system, the server average photon occupation per weight and the client gain $G$ (see Fig.~\ref{Fig_supp_map} in the Methods section). Recalling that the weight leakage and the data leakage are both functions of the hardware configuration parameters, we perform a change of variables and replace the axes in Fig.~\ref{Fig_supp_map} with weight and data leakage to obtain the tradeoff result in Fig.~\ref{Fig_main_maps}. Note that the data leakage is calculated for the quantum adversary $(k=2)$.

The figure depicts a clear tradeoff of the data leakage and weight leakage in order to maintain classification accuracy. This is because, when the server sends less energy per weight in order to limit weight leakage, the client has to increase the gain to preserve accuracy of the computation, thereby exposing more of its data. The data leakage increases monotonically with the gain and saturates, achieving the measure and feed-forward leakage. Therefore, as the server laser power is reduced and the client gain is increased, one moves from right to left on the accuracy isocontours in Fig.~\ref{Fig_main_maps}. For example, the 96\% classification accuracy isocontour shows a weight leakage of less than $I_{w_i}=0.1$ bits per symbol and data leakage of less than $I_{x_i}=0.01$ bits per symbol. 

Low-precision inference is extensively studied in academia and industry, with research indicating that reliable inference typically requires at least 8 bits per weight, or as few as 1 bit using state-of-the-art quantization techniques \cite{gholami2022survey}. This requirement is an order of magnitude higher than the upper bound on weight leakage for our protocol. Furthermore, assuming that a dishonest client does not have access to the training dataset, they would be unable to use the leaked information to infer the rest of the model through training.



\begin{figure}[ht!] 
\begin{centering}
\includegraphics[width=\columnwidth]{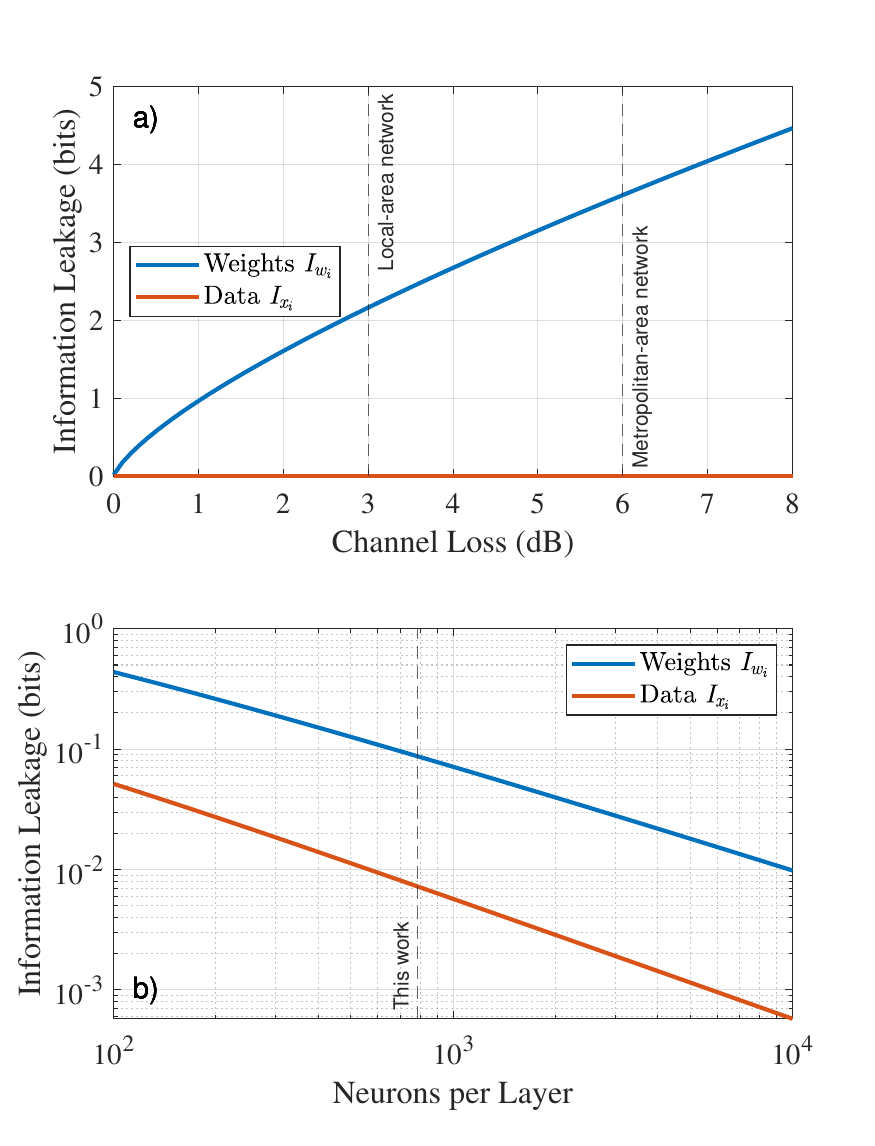}
\par\end{centering}
    \caption{ \textbf{ Information leakage versus channel loss and neurons per layer.} (a) In the presence of loss, the server can increase the average photon occupation per weight to conserve classification accuracy. This increases the weights leakage while the data leakage remains unaffected. The client cannot compensate for high losses, as the classification accuracy saturates as a function of the gain. A weight leakage of up to $4$ bits per weight is obtained for standard losses in local-area networks (LAN) and metropolitan-area networks (MAN). (b) Both weight and data leakages diminish with an increasing number of neurons per layer. Therefore, advanced DNNs, which are much larger than the model used in this work, are expected to suffer from even smaller leakages.}
    \label{Fig_3}
 \end{figure}

The optical loss present between the different parties in real-life communication networks affects the leakage calculations we present above. We analyze the effect of loss on security in Section \ref{Methods} and show that weight leakage increases with round-trip channel loss, while data leakage is independent of round-trip channel loss; the loss-adjusted weight and data leakage are presented in Fig.~\ref{Fig_3}(a). Here we chose an average photon number of $\mu=4$ and amplification gain of $G=3$, the values we used to yield a MNIST classification accuracy of $96\%$. For local area networks or metropolitan areas networks with fiber losses of up to $6$ dB \cite{agrawal2012fiber}, the model leakage is at most $4$ bits per weight. 

Since the number of neurons per layer of state-of-the-art DNNs is much larger than the size presented in our study, we present the dependence of the data and weight leakages on the number of neurons per layer in Fig. ~\ref{Fig_3}(b). As in pannel (a), we chose an average photon number of $\mu=4$ and an amplification gain of $G=3$, the values we used to yield a MNIST classification accuracy of $96\%$.  Interestingly, the leakage vanishes with the number of neurons per layer. This is because the amplification excess noise is spread over more modes in the verification state, leading to less weight and data leakages. Therefore, advanced DNNs, which are much larger than the model used in this work, are expected to suffer even smaller information leakage.

Finally, we briefly introduce an extension of our protocol to a network consisting of a server and two or more clients, where each party assumes that all other parties are dishonest and collaborate against them. We propose two network architectures: symmetric and asymmetric. 

In the symmetric network, the server splits the light encoding of the weights in half and sends each portion to a different client. The clients use the same gain for inference and return verification states to the server. The server coherently combines the verification states and measures the noise in the output. The added excess noise is doubled relative to a single client, resulting in an increase in weight leakage (Eq. \eqref{eq:wleakage}). However, the data leakage for each client does not change compared to a single client (Eq. \eqref{eq:xleakage}). Since the light received by each client is halved in power, the SNR drops by a factor of two for each client, leading to a reduction in classification accuracy. In the asymmetric scheme, client 1 uses all the light for inference and transmits its entire verification state to client 2. Client 2, after performing its own inference, sends the doubly disturbed verification state back to the server, which measures the excess noise. The advantage of the asymmetric architecture is that it eliminates the need for coherent combining of the verification states from both clients at the server. 

\section{Discussion}

Our current protocol forces a trade-off between security for classification accuracy; future generations could relax this trade-off by joint optimization of DNN model parameters and protocol steps. Additionally, better performance may be achieved in a hardware-aware environment by optimizing the weights via physical custom layers instead of noiseless layers. Performance may be further improved by incorporating non-Gaussian operations into the protocol \cite{xia2021quantum,PhysRevX.9.041023}. For example, the client could use a photon-number-resolving detector, and the server could encode the weights using the photon number. The use of quantum resources will also enable the extension of our protocol beyond classical computation \cite{harrow2009quantum}, such as quantum machine learning \cite{rebentrost2014quantum,skolik2021layerwise, biamonte2017quantum, cong2019quantum} or blind quantum computation \cite{broadbent2009universal,barz2012demonstration}. 

Our security analysis for weight and data leakage assumes individual attacks, continuous modulation of the encoding state, and asymptotically large blocks of signals per communication. Recently, however, asymptotic and finite-size security analyses of the discrete modulated continuous-variable quantum key distribution have been introduced for coherent attacks \cite{lin2019asymptotic, kanitschar2023finite,ghorai2019asymptotic}. These results could be applied to our quantum-secure multiparty deep learning to achieve security against other attack classes by adapting them to optimize for the security metric of leakage in our protocol.

Recent experimental advances support the feasibility of our system. On the client side, large unitary transformations and optical computations have recently been demonstrated \cite{bogaerts2020programmable, wang2022optical, polino2019experimental, saggio2021experimental, agresti2019pattern, marpaung2019integrated}. On the server side, continuous-variable quantum key distribution has been successfully implemented \cite{jouguet2013experimental, jain2022practical, zhang2020long}. The combination of these technologies paves the way for the near-term realization of our quantum secure multiparty deep learning.


In this work, we focus on secure inference; however, the protocol can also be applied to secure training. For example, our coherent linear algebra engine (Section \ref{Sec_engine}) can guarantee a secure inner product for federated learning, where multiple clients collaborate to train a model while ensuring that their data remain secure.


Our approach solves a fundamental security challenge in multiparty computation, paving the way for information-theoretical security at various stages of the machine-learning pipeline. The introduction of our protocol enables the secure application of deep learning on private datasets and models, establishing a rigorous standard of secure computation in sectors such as finance, healthcare, and business.

\section{Methods} \label{Methods}

\subsection{Optical implementation} \label{Methods_opt}

\begin{figure}[!ht]
\begin{centering}
\includegraphics[width=\columnwidth]{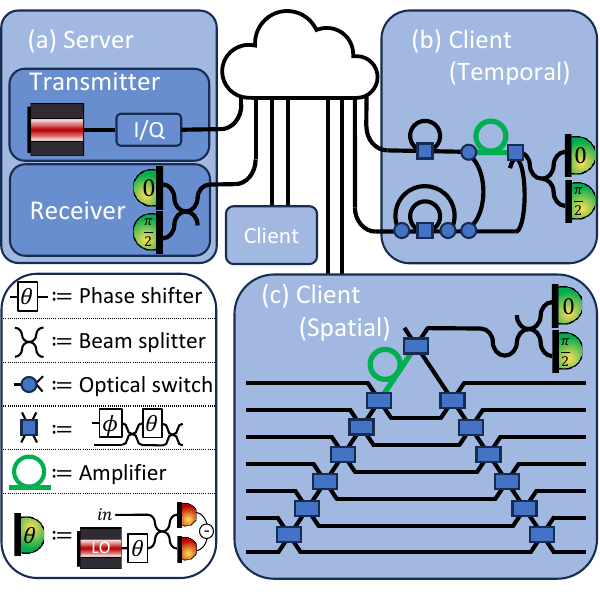}
\par\end{centering}
\caption{\textbf{Optical implementation.} (a) The server transmits neural network weights, encoded by complex amplitudes of weak coherent states generated by an attenuated laser. The encoding can be either temporal or spatial. (b, c) The client performs a series of linear operations on the beam to measure the inner product of a local data vector with the distributed vector (see main text). After these operations, the client returns the residual light as a verification state to the server, which measures the excess noise to compute the leakage of its weights. In the spatial domain (c), the client uses an MZI mesh to interfere spatially adjacent modes, resulting in the uppermost output mode having a complex amplitude representing the desired inner product. This mode is then amplified (green loop) and split by a weighted beam splitter (top) into a coherent homodyne detector (green detectors) and into another MZI mesh. The second MZI mesh implements the inverse of the first mesh. In the time domain (b), the first MZI mesh is replaced with a single MZI and a fiber loop, while the second MZI mesh is replaced by nested fiber loops.}
 \label{Fig_opt}
 \end{figure}

The server consists of two modules, the transmitter and the receiver (see Fig. \ref{Fig_opt}(a)). The transmitter module I/Q modulates the neural network weights onto a train of weak coherent states which are produced by attenuating a continuous-wave laser to the few-photon limit. The receiver module measures the modulated quadratures of the incoming verification state using homodyne detection with a reference local oscillator. The optical power difference between the two output arms is proportional to either the I or the Q quadrature of the verification state, depending on the phase of the local oscillator.

We propose multiple optical implementations of the unitaries $U_{\hat{x}}$ and $U_{\hat{x}}^{\dagger}$ in the client. The client could operate via time domain encoding using optical loops \cite{motes2014scalable} (see Fig. \ref{Fig_opt}(b)). Equivalently, the client could operate by spatial domain encoding using a mesh of interferometers \cite{reck1994experimental,bogaerts2020programmable} (see Fig. \ref{Fig_opt}(c)) or free-space multi-plane light converters \cite{fontaine2019laguerre, lib2022processing, lib2024high}. Time-domain implementations are promising for large-scale universal optical information processing because the number of elements does not scale with the number of optical modes \cite{he2017time, arrazola2021quantum, yonezu2023time}. However, because of optical loss, both technologies are currently limited to a few dozen modes. 

In the time domain design, optical modes are defined by temporally separate pulses. The first operation $U_{\hat{x}}$ is implemented using a Mach-Zehnder interferometer (MZI) and a fiber loop. This structure allows us to interfere successive pulses from the incoming pulse train $\vec w$ with a pulse in the loop that contains the running sum of the inner product, resulting in the final inner product $\vec{w}\cdot \hat{x}$ being written into the complex amplitude of the last output pulse. The light in this mode is then amplified by a factor of $G$ using a phase-insensitive amplifier such as a standard erbium-doped fiber amplifier. Then a beam splitter divides the light with a splitting ratio of $1-\frac{1}{G}:\frac{1}{G}$ for the transmitted and reflected ports. The transmitted port is then measured using homodyne detection in both quadratures to estimate the desired inner product. The reflected port is fed into a nested Mach-Zehnder fiber loop that implements the unitary $U^{\dagger}_{\hat x}$ and routes the resultant verification state back to the server.

In the spatial domain, optical modes are defined by spatially separated waveguides. Here, the client implements the same $U_{\hat x}$ and $U^\dagger_{\hat x}$ via two series of MZIs weighted according to the elements of $\hat x$ \cite{reck1994experimental}. These MZIs mix spatially adjacent modes to shift the inner product into the top mode, which is then amplified and measured via coherent homodyne detection as before.

\subsubsection{Quantum description}

\subsubsubsection{Phase-insensitive amplifier}
We use the quantum electromagnetic field ladder operators (or annihilation and creation operators), denoted by \( \hat{a} \) and \( \hat{a}^\dagger \), respectively, to analyze the behavior of the system at the quantum level. The action of these operators on the photon number states \( |n\rangle \) is given by:
\begin{equation*}
\hat{a} |n\rangle = \sqrt{n} |n-1\rangle \quad \text{and} \quad \hat{a}^\dagger |n\rangle = \sqrt{n+1} |n+1\rangle
\end{equation*}

The input-output relationship of the ladder operators in the Heisenberg picture for a phase-insensitive amplifier with gain \( G \) is \cite{scully1997quantum}:
\[
\hat{a}_{\text{amp}} = \sqrt{G} \hat{a}_{\text{in}} + \sqrt{G-1} \hat{a}_{\text{v}}^\dagger
\]
where \( \hat{a}_{\text{in}} \) is the annihilation operator for the input mode, \( \hat{a}_{\text{amp}} \) is the annihilation operator for the amplified output mode, and \( \hat{a}_{\text{v}} \) is the annihilation operator for the auxiliary vacuum noise mode introduced during amplification.

The quadrature operators are defined as:
\begin{equation*}
\hat{X} =  \hat{a} + \hat{a}^\dagger, \quad \hat{P} =  \frac{1}{i}(\hat{a} - \hat{a}^\dagger)
\end{equation*}
The variance for an operator \( \hat{O} \) is given by:
\[
\langle (\Delta \hat{O})^2 \rangle = \langle \hat{O}^2 \rangle - \langle \hat{O} \rangle^2
\]
Then, the following is true for the input mode:
\[
\langle (\Delta \hat{X}_{\text{in}})^2 \rangle = 1,\quad
\langle (\Delta \hat{P}_{\text{in}})^2 \rangle = 1
\]
For the output mode of the phase-insensitive amplifier:
\begin{equation*}
\hat{X}_{\text{amp}} = \hat{a}_{\text{amp}} + \hat{a}_{\text{amp}}^\dagger, \quad \hat{P}_{\text{amp}} = \frac{1}{i}(\hat{a}_{\text{amp}} - \hat{a}_{\text{amp}}^\dagger)
\end{equation*}
Substituting the expression for \( \hat{a}_{\text{amp}} \):
\begin{align*}
\hat{X}_{\text{amp}} &= \sqrt{G} \hat{X}_{\text{in}} + \sqrt{G-1} \hat{X}_{\text{v}}\\
\hat{P}_{\text{amp}} &= \sqrt{G} \hat{P}_{\text{in}} - \sqrt{G-1} \hat{P}_{\text{v}}
\end{align*}
The variance of the quadratures after amplification is:
\begin{align*}
\langle (\Delta \hat{X}_{\text{amp}})^2 \rangle &= \langle (\sqrt{G} \hat{X}_{\text{in}} + \sqrt{G-1} \hat{X}_{\text{v}})^2 \rangle = 2G - 1\\
\langle (\Delta \hat{P}_{\text{amp}})^2 \rangle &= 2G - 1
\end{align*}
This shows that the noise is amplified along with the signal, and for large \( G \), the noise variance scales linearly with \( G \).

\subsubsubsection{Weighted beamsplitter}

The beam splitter has a splitting ratio of $\left( 1-\frac{1}{G} \colon \frac{1}{G}\right)$. In the spatial domain picture, the input modes for the beam splitter are the amplified output $\hat a_{\text{amp}}$ and a vacuum mode, labeled $\hat b_{\text{vac}}$. The weighted beamsplitter transform yields:
\begin{align*}
\hat a_{\text{out1}} &= \sqrt{1 - \frac{1}{G}} \hat a_{\text{amp}} - \sqrt{\frac{1}{G}} \hat b_{\text{vac}}\\
\hat a_{\text{out2}}&= \sqrt{\frac{1}{G}} \hat a_{\text{amp}} + \sqrt{1 - \frac{1}{G}} \hat b_{\text{vac}}
\end{align*}
Substituting the above expressions for $\hat X_{\text{amp}}, \hat P_{\text{amp}}$:
\[
\hat X_{\text{out2}} = \sqrt{\frac{1}{G}} \hat X_{\text{amp}} + \sqrt{1 - \frac{1}{G}} \hat X_{\text{vac}}
\]
and similarly for $\hat P_{\text{out2}}$.

Since \( \hat{X}_{\text{vac}} \) is the quadrature of the vacuum mode, it has unit variance: $ \langle (\Delta \hat{X}_{\text{vac}})^2 \rangle = 1 $.

Notice that two vacuum modes have been introduced: one during amplification to preserve the commutation relation of the amplified output mode, and one at the dark input port of the beamsplitter. The two vacuum modes are independent and uncorrelated, so they can be summed in quadrature. One can see this directly by considering the reparametrization:
$\sqrt{G} \mapsto \cosh(\theta)$ and $\sqrt{G-1} \mapsto \sinh(\theta)$ (note that $\cosh^2(x) = \sinh^2(x) + 1$ for any $x$). Their addition in quadrature forms a single vacuum noise mode with unit variance with a coefficient determined by the Pythagorean theorem sum of their composite variances.

Alternatively, we can derive the noise variance directly as follows:
\begin{align*} \langle (\Delta \hat{X}_{\text{out1}})^2 \rangle  = \langle (\Delta \hat{P}_{\text{out1}})^2 \rangle = 1 + \left(2 - \frac{2}{G} \right)
\end{align*}
The excess noise on the amplified signal after beam splitting is thus $\left(2 - \frac{2}{G} \right)$.

For \( \hat{a}_{\text{out2}} \):
\[ \hat{X}_{\text{out2}} = \sqrt{1 - \frac{1}{G}} \hat{X}_{\text{amp}} - \sqrt{\frac{1}{G}} \hat{X}_{\text{vac}} \]

and similarly for $\hat P_{\text{out2}}$.
Thus,

\begin{align*}
\langle (\Delta \hat{X}_{\text{out2}})^2 \rangle = \langle (\Delta \hat{P}_{\text{out2}})^2 \rangle = & \frac{2G^2-3G+2}{G} \end{align*}

The optical power at the client's detector, after gain and splitting, is $S = G(1-\frac{1}{G})|\vec{w}\cdot \hat{x}|^2$. The SNR of the client's measurement is therefore: 

\begin{equation} \label{eq_SNR_supp}
\text{SNR}= \frac{G(G-1)}{2G^2-3G+2} |\vec{w}\cdot \hat{x}|^2 
\end{equation}

\subsubsubsection{Verification state excess noise} \label{Met_excess}

First, we consider the effect of the overall unitary transform pair $\{ U_{\hat x}, U_{\hat x}^\dagger \}$ on the variance of the state. The initial input $\vec w$ is a set of coherent states with a unit shot-noise variance in each mode, that is, the noise itself is uncorrelated and isotropic. We describe the noise in these $N$ modes and the correlations between them, through the covariance matrix of their quadratures $\Sigma^X$. The $(i,j)$-th element of this matrix is given by $\Sigma^X{(i,j)}~=~\mathbb{E}\left[X_iX_j\right]-\mathbb{E}\left[X_i\right]\mathbb{E}\left[X_j\right]$.

Since a coherent state is equivalent to a displacement operation on the vacuum state, the covariance matrix of $\vec w$ is $\Sigma^X_{w} = \mathbb{1}$. Since we want the first output mode (the result mode) of $U_{\hat x}\vec w$ to carry the inner product $\vec w\cdot \hat x$, $U_{\hat x}$ and $U_{\hat x}^\dagger$ take the form:
\[ U_{\hat x} \sim
\mleft[
\begin{array}{ccc}
\multicolumn{3}{c}{\hat{x}} \\ \hline
 &  & \\
 & \mathbf{v} & \\
 &  &
\end{array}
\mright]\; ; \;
U_{\hat x}^\dagger \sim 
\mleft[
\begin{array}{c|ccc}
& & & \\
\hat{x}^\dagger & & \mathbf{v}^\dagger & \\
& & &
\end{array}
\mright]
\]
where $\hat x$ is represented as a row vector and $\mathbf{v}$ is a (non-unique) rectangular matrix which is chosen to ensure unitarity of $U_{\hat x}$.

The covariance matrix of $U_{\hat x}\vec w$ is:
\[
U_{\hat x} \Sigma^X_{w} U^\dagger_{\hat x} = U_{\hat x} U^\dagger_{\hat x} = \mathbb{1}
\]
Next, we apply the amplify-and-split operation $\mathcal{G}(\cdot)$ on the vector of modes $U_{\hat x}\vec w$. This operation involves applying a phase-insensitive amplification with gain $G$ on the result mode and then splitting it with the ratio $1-\frac{1}{G}:\frac{1}{G}$. The output of the $\frac{1}{G}$ port is fed back to the mesh that implements $U_{\hat x}^\dagger$. The operation $\mathcal{G}(\cdot)$ leaves all the other modes of $U_{\hat x}\vec w$ untouched. The resultant covariance matrix after amplification and splitting is the diagonal matrix $\Sigma^X_{\mathcal{G}(U_{\hat{x}}\vec{w})} \sim \operatorname{diag}(\sigma^2, 1, 1, \ldots, 1)$, where $\sigma^2 = 1 + \left(2 - \frac{2}{G} \right)$ is the variance of the result mode calculated in Section \ref{Methods_opt}. 

Finally, we operate with the transform $U^\dagger_{\hat x}$:
\begin{align*}
\Sigma^X_{U_{\hat{x}}^{\dagger}\mathcal{G}(U_{\hat{x}}\vec{w})} &= U^\dagger_{\hat x} \Sigma^X_{\mathcal{G}(U_{\hat{x}}\vec{w})} U_{\hat x} = \mathbb{1} + \hat{x}^{\dagger} \hat{x} \left(2 - \frac{2}{G} \right)
\end{align*}
Note that since $\hat{x}^{\dagger}\hat x$ is a dense matrix, the various components of the verification state $U_{\hat{x}}^{\dagger}\mathcal{G}(U_{\hat{x}}\vec{w})$ are all correlated with one another. However, the variance of the $i$-th output mode only contains information from $\hat{x}_i$; recalling that we had defined the excess variance of the $i$-th output mode as the excess noise $\eta_i$ in Section~\ref{Sec_engine}, we have: 
\[
\eta_i = \left(2 - \frac{2}{G} \right) \cdot \abs{\hat{x}_i}^2
\]

\subsection{Classification accuracy calculation} \label{Methods_class}
To analyze our protocol's performance for deep learning, we compute the classification accuracy for the standard MNIST classification task using PyTorch. To this end, we write custom neural network layers that implement our coherent linear algebra engine presented in Section \ref{Sec_engine}; several of these custom layers were strung together to form a secure neural network. The homodyne current detected in each custom layer is sampled from a unit Gaussian distribution $\mathcal{N}(0,1\text{ SNU})$ to account for the quantum shot noise in each quadrature. The resultant noisy sample is then fed to the non-linear ReLU activation function before being passed on to the next custom layer. 

Separately, we trained a digital model composed of standard PyTorch layers on preprocessed MNIST images. The preprocessing of the data set included the flattening of the $28\times28$ images to vectors of size 784, followed by a centering and scaling of each vector to the range [-1, 1]. Then we trained a standard 2-layer network with 784 inputs, 784 hidden neurons, and 10 outputs, obtaining a classification accuracy of $98\%$. We then transferred these trained digital weights into the custom secure neural network and computed the accuracy achieved by the secure protocol as a function of the weight pulse energy. We found that the accuracy of the digital and analog models were in agreement up to variations caused by quantum shot noise. 

The real operations required for standard deep learning tasks can be implemented on our complex-valued hardware by encoding the real input vector $\vec x^{(\mathbb{R})}$ of length $N$ and the real weight matrix $W^{(\mathbb{R})}$ of size $M\times N$ into a complex input vector $\vec x^{(\mathbb{C})}$ of length $N/2$ and a complex weight matrix $W^{(\mathbb{C})}$ of shape $M\times N/2$ using the following procedure: 
\begin{equation*}
W_{i,k}^{(\mathbb{C})} = W_{i,2k-1}^{(\mathbb{R})} + iW_{i, 2k}^{(\mathbb{R})} \quad \text{and} \quad x_k^{(\mathbb{C})} = x_{2k-1}^{(\mathbb{R})} - ix_{2k}^{(\mathbb{R})}.
\end{equation*}
The hardware computes the real value of the product $W^{(\mathbb{C})}x^{(\mathbb{C})}$, which is precisely the desired matrix-vector product $W^{(\mathbb{R})}x^{(\mathbb{R})}$ required for the DNN computations.

We define the signal-to-noise ratio (SNR) at the homodyne detector at the neural network outputs as 

\begin{equation*}
\text{SNR}_i= \frac{G(G-1)}{2G^2-3G+2} |\vec{w} \cdot \hat{x}|^2
\end{equation*}
where The amplitudes of these physical states, \(w_j\), encode the logical weights as \(w_j = \sqrt{\mu} \mathbf{W}_{ij} / \|\mathbf{W}\|_{\text{RMS}}\), where \(\|\mathbf{W}\|_{\text{RMS}}\) denotes the root mean square of the elements in \(\mathbf{W}\). Therfore we have:

\begin{equation*}
\text{SNR}_i~=~\frac{G(G-1)}{2G^2-3G+2}\frac{\mu}{\|\mathbf{W}\|^2_{\text{RMS}}} |\mathbf{W}_{i} \cdot \hat{x}|^2
\end{equation*}

Note that in order to obtain the desired inner product, the client scales the result digitally by $\frac{\|\vec x\|}{\sqrt{G-1}}\frac{\|\mathbf{W}\|_{\text{RMS}}}{\sqrt{\mu}}$. However, for the purpose of deep learning, the scaling step for the client is superfluous because the client normalizes between layers regardless. 
We define a "physical scaling parameter" \(F\) to capture the hardware-dependent prefactor in the SNR equation,

\begin{equation*}
F := \sqrt{\frac{G(G-1)}{2G^2-3G+2} \mu}
\end{equation*}

\begin{figure}[ht!] 
\begin{centering}
\includegraphics[width=\columnwidth]{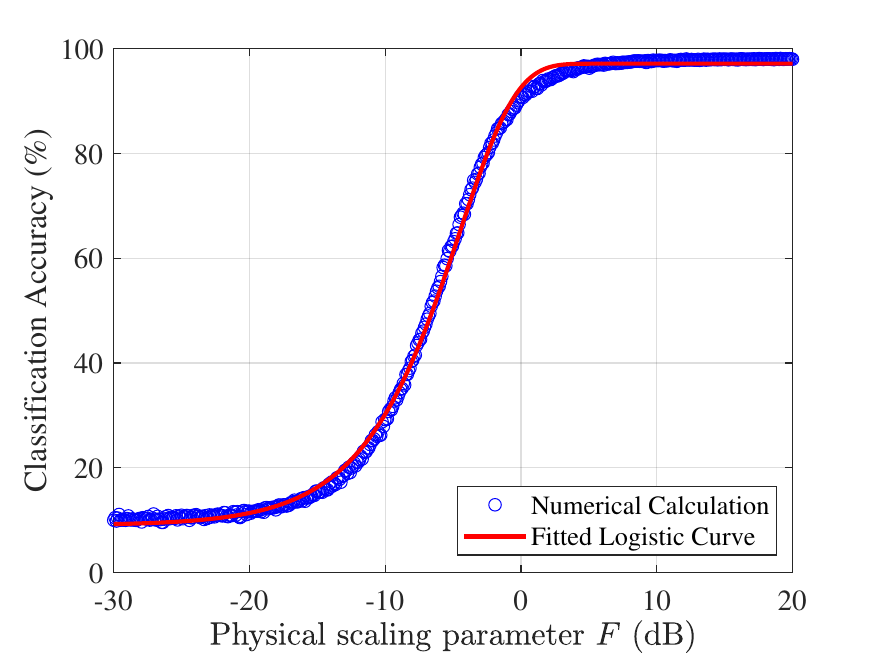}
\par\end{centering}
    \caption{ \textbf{Classification accuracy versus the physical scaling parameter $F$.}  We calculate the classification accuracy considering an additive noise to each inner product, using weights trained for a digital noiseless model. We present the classification accuracy as a function of the the physical scaling parameter $F$. The classification accuracy increases monotonically with $F$, asymptotically obtaining the digital noiseless accuracy. This parameter capture the hardware-dependent prefactor in the SNR, allowing to calculate the the mutual effect of the average photon number $\mu$ and the amplification gain $G$. We fit the classification accuracy using a logistic curve. } \label{Fig_supp_acc}
 \end{figure}

 Varying \(F\) changes the SNR of the inner product outputs, thus affecting the accuracy of the secure neural network. The purpose of defining this parameter is to enable the calculation of the classification accuracy as a function of the gain \(G\) and average photon occupation \(\mu\) without directly executing the intensive numerical calculations in the two-dimensional space spanned by these two parameters. Instead, we calculate the classification accuracy within the space spanned by \(F\). The dependency of the test accuracy achieved by the secure neural network on $F$ is presented in Fig.~\ref{Fig_supp_acc}. We further find that a logistic function faithfully fits the variation of the classification accuracy as a function of $F$:

\begin{equation*}
    \text{Acc}(F) = \frac{L}{1 + e^{-k(F-F_0)}} + B
\end{equation*}
with Root-Mean-Square-Error (RMSE) of $1.1\%$. This fit allows us to use a convenient analytical formula to predict the classification accuracy as a function of the hardware parameters $G$ and $\mu$, instead of using look-up tables. 

\begin{figure}[ht!] 
\begin{centering}
\includegraphics[width=\columnwidth]{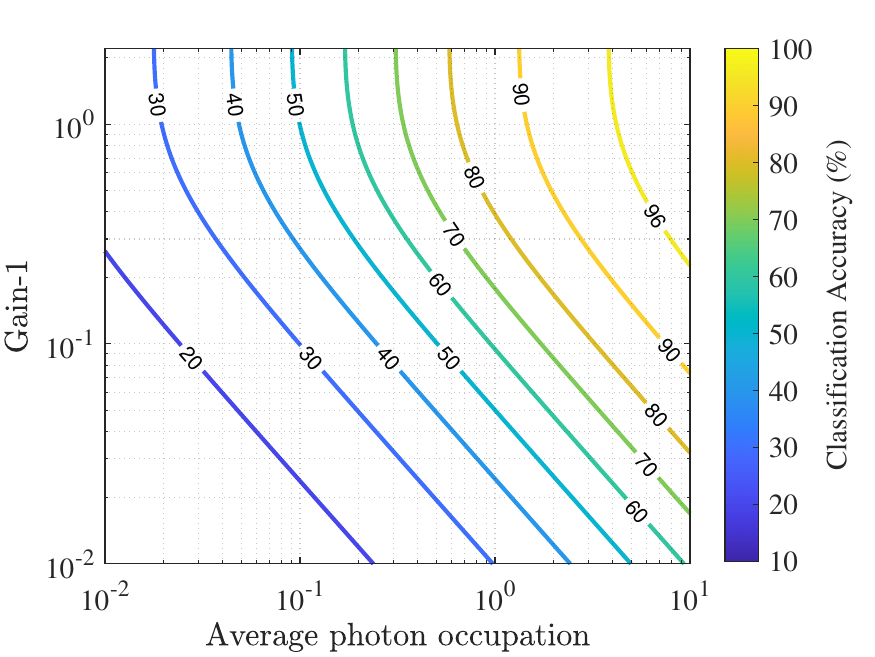}
\par\end{centering}
\caption{\textbf{Classification accuracy versus amplification gain and data leakage.} The classification accuracy of the physical model, which uses our coherent linear algebra engine (Fig.~\ref{Fig_Ilus}), is numerically calculated as a function of the average photon occupation per weight and the amplification gain at the client. The classification accuracy increases with gain and average occupation number and asymptotically achieves the digital noiseless accuracy. At low gain values, the gain is inversely proportional to the photon occupation number for any given fixed classification accuracy. }
\label{Fig_supp_map}
 \end{figure}

We present the classification accuracy as a function of the average photon occupation per weight and the amplification gain at the client in Fig. \ref{Fig_supp_map}. The classification accuracy increases with gain and average occupation number and asymptotically achieves the digital noiseless accuracy. At low gain values, the gain required to maintain a fixed classification accuracy is inversely proportional to the average photon occupation number per weight.

\subsection{Detailed security analyses} \label{Methods_sec}

We analyze the security of this protocol following the standard security analysis for continuous-variable quantum key distribution (CVQKD).

First, we express the setting of our problem in the continuous-variable quantum key distribution (CVQKD) framework. In a standard CVQKD setting, two legitimate parties Alice and Bob wish to communicate a message under possible attack by a third-party eavesdropper Eve external to both of them. In the CVQKD protocol, normal operation results in Bob receiving the message from Alice, and the mutual information between Alice and Eve (along with Eve's Holevo information) must be bounded below an arbitrary threshold.

In the context of our multiparty quantum-secured DNN protocol, normal operation results in the transfer of a specific amount of information from the server to the client \textemdash\ specifically, the inner product of the server's weights with the client's data must be extracted by the client in order to satisfy the accuracy requirements. Here, both parties (server and client) have the opportunity to act maliciously.

When the server is honest and the client is malicious, the information that is at risk of leakage is the weight vector sent from the server to the client. When the server is malicious and the client is honest, the information that is at risk of leakage is the client data vector used for the client's inner product. We only seek to protect the information of the honest party, so the case in which both parties act maliciously is not relevant to our protocol.

The eavesdropper may operate under different models based on different capabilities; here, we restrict our analysis to individual attacks. Under individual attacks, Eve performs independent and identically distributed (i.i.d.) attacks on all incoming modes; that is, they prepare separable ancilla states that interact individually with one of the signal modes in the quantum channel. The ancilla states are then stored in a quantum memory until measurement and are measured independently of each other. Extensions to collective and coherent attacks are left for future work.

\subsubsection{Weights Leakage}
\label{security_weights}

From the perspective of a malicious client, the server prepares displaced coherent states with quadrature components \( q \) and \( p \) that are realizations of two i.i.d. random variables \( Q \) and \( P \). These random variables are normally distributed as \( Q, P \sim \mathcal{N}(0, V_{\text{mod}}) \), where \( V_{\text{mod}} \) denotes the modulation variance.

Here we analyze the security against individual attacks, defined above, on the incoming modes of $\vec w$ sent by the honest server.

Using the source replacement method commonly applied in QKD security analyses, the state sent by the server can be described as an entangled state. In this scenario, the server prepares a two-mode squeezed vacuum state (TMSVS), measures both quadratures of one mode, and sends the other mode to the client. In shot-noise units, the TMSVS is characterized by the covariance matrix \cite{laudenbach2018continuous}:

\begin{equation*}
\Sigma_{\text{TMSVS}} = \begin{pmatrix}
V\mathbb{1} & \sqrt{V^2-1}\sigma_z  \\
\sqrt{V^2-1}\sigma_z & V\mathbb{1} 
\end{pmatrix}
\end{equation*}

Here, \( \sigma_z \) is the Pauli matrix, and \( V \) represents the variance of the quadrature operators. The upper and lower diagonal $2\times 2$ blocks of $\Sigma_{\text{TMSVS}}$ are the covariance matrices of the quadrature operators of Alice and Bob respectively, while the off-diagonal block matrices are the covariances between the quadratures of Alice and those of Bob. The variance \( V \) is related to the mean photon number per pulse by the relation \(\mu = \frac{1}{2}(V-1)\). This variance is the sum of Alice's actual modulation variance of her two quadrature components and the vacuum shot-noise variance.

After Bob’s mode is transmitted through the channel with transmission coefficient \( T \) and excess noise \( \xi \), the covariance matrix transforms to:

\begin{equation*} \label{cov}
\Sigma_{AB} = \begin{pmatrix}
V\mathbb{1}& \sqrt{T(V^2-1)}\sigma_z  \\
\sqrt{T(V^2-1)}\sigma_z & (T(V-1) + 1 + \xi) \mathbb{1}
\end{pmatrix}
\end{equation*}

Without loss of generality, we consider \( \ket{\psi} \) as a bipartite state with Alice’s and Bob’s joint subsystem comprising one part and Eve’s subsystem the other. Any pure bipartite state can be expressed using the Schmidt decomposition as \( \ket{\psi} = \sum_i \lambda_i \ket{i}_{AB} \ket{i}_E \). Tracing out either subsystem yields a mixed state with a von Neumann entropy that depends only on the Schmidt coefficients \( \lambda_i \):

\begin{equation*}
S_E = S_{AB} = -\sum_i \lambda_i \log(\lambda_i)
\end{equation*}

We use the Holevo theorem to bound the amount of information that Eve can obtain \cite{holevo1973bounds}. 

The Holevo theorem provides an upper bound on the accessible information that can be extracted from a quantum system. Tracing out Eve's subspace from the joint state $\ket{\psi}\bra{\psi}$ yields Alice's and Bob's joint substate, that is, $\mathrm{Tr}_E(\ket{\psi}\bra{\psi}) = \rho_{AB}$. Mathematically, the Holevo bound \( \chi \) for the information held by Eve's substate $\rho_E = \mathrm{Tr}_{AB}(\ket{\psi}\bra{\psi})$ about the quantum state \( \rho_{AB} \) shared between Alice and Bob is given by \cite{laudenbach2018continuous}:

\begin{equation*}
\chi  = S_E - S_{E|A} =  S_{AB} - S_{B|A}
\end{equation*}
where $S_{X|Y}$ is the conditional von Neumann entropy of the density matrix $\rho_X$ given the density matrix $\rho_Y$. 

The von Neumann entropy of a quantum state can be expressed in terms of the \textit{symplectic eigenvalues} of that state's covariance matrix. The symplectic eigenvalues of a \(2N \times 2N\) matrix \(\Sigma\) are defined as the absolute values of the ordinary eigenvalues of the matrix $\bar{\Sigma} = i \Omega \Sigma $ where \(\Omega\) is the \(2N \times 2N\) matrix given by the direct sum:
\[
\Omega = \bigoplus_{l=1}^{N} \begin{pmatrix}
0 & 1 \\
-1 & 0
\end{pmatrix}
\]
It follows that a Gaussian state with covariance matrix \(\Sigma\) has a von Neumann entropy given by $ S = \sum_i g(\nu_i) $, where

\begin{equation*}
g(\nu) = \left( \frac{\nu + 1}{2} \right) \log_2 \left( \frac{\nu + 1}{2} \right) - \left( \frac{\nu - 1}{2} \right) \log_2 \left( \frac{\nu - 1}{2} \right) \quad 
\end{equation*}
and \(\nu_i\) is the $i$-th symplectic eigenvalue of \(\Sigma\) (see \cite{laudenbach2018continuous} for details).

To calculate $S_{AB}$ we calculate the symplectic eigenvalues of $\Sigma_{AB}$:

\begin{equation*}
\nu_{1,2} = \frac{1}{2} \left( z \pm \left[ b - a \right] \right) 
\end{equation*}
where
\begin{alignat*}{2}
    &a:= V \quad &&b:=  T(V-1)+1+\xi\\\
    &c:= \sqrt{T(V^2-1)} \quad &&z:= \sqrt{(a + b)^2 - 4c^2},
\end{alignat*}
Then, we have $S_{AB}=g(\nu_1)+g(\nu_2)$. Note that in Section \ref{Sec_security} we presented these quantities using occupations $\mu = \frac{V-1}{2}$ and $T=1$. We analyze the effect of optical loss where the transmission $T<1$ in Section \ref{Sec_secure_classiffication}.

We calculate \( S_{B|A} \) assuming that Bob implements homodyne detection. The covariance matrix $\Sigma_{B|A}$ is given by:
\begin{equation*}
\Sigma_{B|A} = \Sigma_B - \frac{1}{V_A + 1} \Sigma_C \Sigma_C^T 
\end{equation*}
where \(\Sigma_B = b \mathbb{1}\), \(\Sigma_C = c \sigma_z\), \(V_A = a\), and \(a\), \(b\), and \(c\) are defined as above. Using these definitions, we obtain
\begin{equation*}
\Sigma_{B|A} = \begin{pmatrix} 
b & 0 \\ 
0 & b 
\end{pmatrix} 
- \frac{1}{a + 1} 
\begin{pmatrix} 
c^2 & 0 \\ 
0 & c^2 
\end{pmatrix} 
= \left( b - \frac{c^2}{a + 1} \right) \mathbb{1} \quad 
\end{equation*}
Its symplectic eigenvalue is:
\begin{equation*}
\nu_3 = b - \frac{c^2}{a + 1}
\end{equation*}
This gives us $S_{B|A}=g(\nu_3)$, which finally yields $\chi~=~g(\nu_1)+g(\nu_2)-g(\nu_3)$.

\subsubsection{Weight leakage accumulation due to multiple queries}
\label{security_multiquery}
We previously showed that clients only gain limited information on the DNN weights through a single broadcast of the weights. We now discuss how to prevent this information from accumulating during multiple queries. Specifically, we propose to manipulate the weights of the model in every broadcast in a way that preserves the neural network function while drastically reducing the accumulation of weight information at the client' end.
Although neural network functions are not invariant to general linear transformations (e.g. translation \cite{lyle2020benefits,kauderer2017quantifying}), there always exist isomorphisms between weight matrices that do not alter the neural network function. 

Let $W_1,W_2\in \mathbb{R}^{N\times N}$ (this procedure works equally well for rectangular matrices) 
be two weight matrices with $\sigma(\cdot)$ implementing an elementwise nonlinearity,
so $W_2\sigma(W_1)$ implements two successive layers of a neural network. 
We note that for any permutation matrix $\tau$ and its inverse $\tau^{-1}$: $W_2\sigma(W_1)=(W_2 \tau^{-1})\sigma(\tau W_1)$. Thus, there are $N!$ different permutations possible that all return the same computation \cite{NIPS2017_10ce03a1}. The server can send any of these $\tau W$ matrices. This process can be repeated with $W_2$ and the next matrix of the network $W_3$ and so on. For the common case of $\sigma(x)=\mathrm{ReLU}(x)=\max(0,x)$, $\tau$ may also include multiplication by a random scalar \cite{NIPS2015_eaa32c96}
, or even multiplication by a random diagonal matrix, limiting the information in one measurement to the entropy of these random manipulations. By choosing a maximally entropic distribution over the transformation space, Alice can limit Eve's ability to infer weight information between broadcasts.

\subsubsection{Data leakage} \label{security_data}

In this scenario, the client operates legitimately, and the server acts maliciously. Under our eavesdropper model, we assume that the server can perform any arbitrary individual attack upon receiving the verification state $\rho_v$. In particular, the server could be equipped to make quantum measurements, observe quantum correlations, and transmit non-Gaussian states.

We recall from Section \ref{Met_excess} that the excess noise in the $i$-th mode in $\rho_v$ is:

\[
\eta_i = \left(2 - \frac{2}{G} \right) \cdot \abs{x_i}^2
\]

In the following classical leakage analysis, we characterize the leakage of each individual data element $x$ (subscript $i$ suppressed for convenience) under the possibility of an individual classical attack from a malicious server.

Let \( R \) be a Gaussian random variable corresponding to the $i$-th mode of the verification state with mean $w$ and variance \(\sigma^2\), where
\[
\sigma^2 = 1 + \left(2 - \frac{2}{G} \right) x^2
\]
for a known parameter \( G \) and an unknown parameter \( x \). 

We aim to calculate the precision in estimating \( x \) using the Cramér-Rao bound for \( M \) measurements and use this precision to determine the number of bits of information obtained by the server about $x$.

The probability density function (PDF) of \( R \) is given by:
\begin{equation}
\label{PDF}
f(R; x) = \frac{1}{\sqrt{2 \pi \sigma^2}} \exp\left(-\frac{(R-w)^2}{2 \sigma^2}\right)
\end{equation}
where $\sigma^2$ is given above.

The Fisher information \( I_R(x) \) for a single measurement with respect to a given distribution on $R$ is given by:
\[
I_R(x) = \mathbb{E}\left[\left(\frac{\partial}{\partial x} \log f(R; x)\right)^2\right]
\]


Noting that $\frac{\partial \sigma^2}{\partial x} = 2 \frac{(2G - 2)}{G} x$, we compute the derivative of the log-likelihood function:

\[
\frac{\partial}{\partial x} \log f(R; x) = 2 \frac{(2G-2)}{G} x \left( -\frac{1}{2 \sigma^2} + \frac{(R-w)^2}{2 \sigma^4} \right)
\]


Noting that $\sigma$ and $x$ are constants with respect to the distribution over $R$, the Fisher information is
\[
I_R(x) = \left(2 \frac{(2G-2)}{G} x \right)^2 \frac{1}{4 \sigma^4} \mathbb{E}\left[\left( -1 + \frac{(R-w)^2}{\sigma^2} \right)^2\right]
\]


Using the properties of the moments of Gaussian distributions: \(\mathbb{E}[(R-w)^2] = \sigma^2\) and \(\mathbb{E}[(R-w)^4] = 3\sigma^4\), we obtain:
\begin{align*}
I_R(x) &= \left(2 \frac{(2G-2)}{G} x \right)^2 \frac{1}{4 \sigma^4} \left( 1 - 2 \frac{\sigma^2}{\sigma^2} + 3 \right)\\
I_R(x) &= \frac{2 (2 - \frac{2}{G})^2 |x|^2}{\sigma^4}
\end{align*}

For \( M \) independent measurements, the Fisher information is additive:
\[
I_{R,M}(x) = M I_R(x) = \frac{2 M (2-\frac{2}{G})^2 |x|^2}{\sigma^4}
\]

The Cramér-Rao bound for the variance of any unbiased estimator \( \hat{x} \) of $x$ is:
\[
\text{Var}(\hat{x})=\sigma^2_{\hat x} \geq \frac{1}{I_{R, M}(x)} = \frac{\sigma^4}{2 M \left(2-\frac{2}{G}\right)^2 |x|^2}
\]

The precision (standard deviation) for estimating \( x \) is:
\[
\sigma_{\hat x} \geq \sqrt{\frac{\sigma^4}{2 M (2-\frac{2}{G})^2 x^2}} = \frac{\sigma^2}{(2-\frac{2}{G})x \sqrt{2M}}
\]

The number of bits of information \( I_x \) that the estimator $\hat x$ contains about $x$ can be computed by treating $\hat x$ as the received signal when $x$ is sent through an additive Gaussian channel with noise variance $\mathrm{Var}(\hat x)$:
\[
I_x = \frac{1}{2} \log_2\left(1 + \frac{|x|^2}{\sigma_{\hat x}^2}\right) \leq \frac{1}{2} \log_2\left(1 + \frac{8 M (G - 1)^2 |x|^4}{G^2 \sigma^4}\right)
\]
This provides the number of bits of information obtained by the server from \( M \) measurements of the client's verification state.

\subsubsection{Data Leakage under quantum operations}
\label{security_quantum_data}

Although the analysis via classical Fisher information provides an upper bound for the data leakage by lower bounding the variance of an unbiased estimator, a tighter result can be achieved by applying the quantum Cramér-Rao bound (QCRB). To this end, we extend our data leakage analysis to adversarial servers equipped with quantum operations in this section. 

The quantum version of the CRB mirrors the classical construction:
\[
\mathrm{Var}(\hat x) \geq \frac{1}{M\mathcal{F}_R[\rho_x]}
\]
where $\mathcal{F}_R[\rho_x]$ is the quantum Fisher information (QFI) about parameter $x$ under observations of random variable $R$. The QFI of a Gaussian state is \cite{Safrenek_2019}:
\begin{equation}
\label{QFI_def}
\mathcal{F}_R[\rho_x] = \frac{1}{2} \mathrm{Tr} \left[ \Sigma^{-1} \frac{\partial \Sigma}{\partial x} \left( \Sigma^{-1} \frac{\partial \Sigma}{\partial x} \right)^\dagger \right]
\end{equation}
where $\Sigma$ is the covariance matrix of $\rho_x$. In our case, the relevant covariance matrix is that of the quadratures of the $i$-th mode of the verification state $\rho_v$:
\[
\Sigma = \begin{pmatrix}
\sigma^2 & 0 \\
0 & \sigma^2
\end{pmatrix}
\]
where \(\sigma^2 = 1 + \left(2 - \frac{2}{G}\right) x^2\). 

Recalling the computation of $\frac{\partial \sigma^2}{\partial x}$, we find the following derivative of the covariance matrix with respect to the parameter $x$:
\begin{align*}
\frac{\partial \Sigma}{\partial x} = \begin{pmatrix}
\frac{\partial \sigma^2}{\partial x} & 0 \\
0 & \frac{\partial \sigma^2}{\partial x}
\end{pmatrix}= \begin{pmatrix}
2x (2-\frac{2}{G}) & 0 \\
0 & 2x (2-\frac{2}{G})
\end{pmatrix}
\end{align*}
so we have the product:
\begin{align*}
\Sigma^{-1} \frac{\partial \Sigma}{\partial x} = \frac{2x}{\sigma^2}\left( 2-\frac{2}{G}\right) \mathbb{1}
\end{align*}

Substituting the expressions for \(\Sigma^{-1}\) and \(\frac{\partial \Sigma}{\partial x}\) into \eqref{QFI_def} yields the QFI:
\begin{equation}
\label{QFI_explicit}
\mathcal{F}_R[\rho_x]
= \frac{4|x|^2 (2-\frac{2}{G})^2}{\sigma^4}
\end{equation}
which is twice the classical Fisher information. The QCRB then bounds the variance of the estimator as follows:
\begin{equation*}
\text{Var}(\hat{x}) \geq \frac{\sigma^4}{4M |x|^2 (2-\frac{2}{G})^2}
\end{equation*}



The precision and number of bits of information leaked can be calculated similarly to Section~\ref{security_data}. The final result is presented in Eq.~\ref{eq:xleakage}.

Finally, we consider the case where the server transmits non-Gaussian states. Non-Gaussian states are not fully characterized by their covariance matrices, i.e., the inter-mode correlations of non-Gaussian states are not restricted to their first and second order moments. In particular, the client's Gaussian operations on the server's non-Gaussian states also influence the higher order moments, which could further expose the client's data. However, under the assumption of individual attacks, we can make the convenient observation that the effect of Gaussian operations on higher order moments of non-Gaussian states is mediated through the effect on the first and second order moments \cite{meena2023nongauss,olivares2013nongaussops}. By the well-known data processing inequality, this implies that the amount of information contained in measurements of higher order moments is less than or equal to the amount of information in the first two moments. Thus, for individual attacks, the QFI extracted from the higher order moments of a given non-Gaussian state is bounded by the QFI extracted from the Gaussian state with the same first and second order moments. The closed form expression in \eqref{QFI_def} then applies again and the bound in \eqref{QFI_explicit} holds. Thus, our protocol is secure against individual attacks made by a malicious server.

\section{Data availability} The data that support the findings of this study are available from the corresponding author upon reasonable request.

\section{Code availability} The code used in this study is available from the corresponding author upon reasonable request. 

\section{Author information}
\subsection{Contributions} K.S. proposed the protocol, performed the security analyses and numerical simulations, and wrote the manuscript under the guidance of D.E. S.K.V. designed the PyTorch package for optical deep neural networks \ref{Methods_class}. R.H. assisted with the optical realization of the protocol and its quantum description \ref{Methods_opt}. P.I. assisted in the quantum calculation of the Cramér-Rao bound \ref{security_quantum_data}. All authors contributed to the writing.

\section{Ethics declarations}
\subsection{Competing interests} K.S and D.E. have filed a US Patent application 63677972 on quantum-secure multiparty deep learning. The other authors declare no competing interests.

\section{Acknowledgments}
The authors kindly thank Saumil Bandyopadhyay, Leshem Choshen and Rom Dudkiewicz for fruitful discussions and suggestions. 
K.S acknowledges the support of the Israeli Council for Higher Education and the Zuckerman STEM Leadership Program.

\bibliographystyle{unsrt}
\bibliography{refs}
\end{document}